\documentclass[pra,aps,amsfonts,amsmath,superscriptaddress,%
  twocolumn,showpacs,floatfix]{revtex4}
\usepackage{amsfonts}
\usepackage{amsmath}
\usepackage{bm}
\usepackage{epsfig}

\begin{document}

 \title {\bf Detecting bubbles in exotic nuclei}
\author{E. Khan, M. Grasso, J. Margueron, N. Van Giai}
\address{Institut de Physique Nucl\'eaire, 
 Universit\'e Paris-Sud,
IN2P3-CNRS,
F-91406 Orsay Cedex, France}


\begin{abstract}

The occurrence of a bubble, due to an inversion of s$_{1/2}$ state with the
state usually located above, is investigated. Proton bubbles in neutron-rich
Argon isotopes are optimal candidates. Pairing effects which can play
against the bubble formation are evaluated. They cannot prevent bubble
formation in very neutron-rich argon isotopes such as $^{68}$Ar. This pleads
for a measurement of the charge density of neutron-rich argon isotopes in
the forthcoming years, with the advent of electron scattering experiments in
next generation exotic beam facilities such as FAIR or RIBF.

\end{abstract}

\vskip 0.5cm
\pacs{21.10.Ft, 21.10Re, 21.60.Jz, 25.30.Bf, 25.70.De}
\maketitle


\section{Introduction}

The occurrence of bubble nuclei has been studied by Wilson in 1946
\cite{wil46}, in order to describe low-lying excitations as oscillations of
nuclear bubbles. The invoked reason for the bubble existence was an
overestimated ``saturation of the nuclear forces''. Bethe and Siemens have
also investigated this possibility in the sixties \cite{sie67}. The
stability of the bubble was studied using the liquid drop model and this
analysis showed that bubbles were not leading to the minimum energy
configuration, compared to prolate shape with uniform density. During the
seventies and eighties, bubble nuclei were also investigated, still with
liquid drop based models \cite{swi83} or Thomas-Fermi approaches
\cite{boh76}. More recently bubbles were predicted in hyperheavy nuclei
\cite{dec03} with Z $\ge$ 120. In this case, self-consistent
Hartree-Fock-Bogoliubov (HFB) calculations have shown that the ground state
corresponds to a bubble configuration. This effect is due to an interplay
between the Coulomb interaction and the nucleon-nucleon interaction in those
very heavy systems. The optimum spatial position of the bubble has also been
investigated \cite{yu00}. However the existence of such heavy nuclei is
speculative, and bubbles have not been detected yet. It should be noted that
bubble effects may be a general feature in nuclei: they have also been
predicted in superheavy nuclei around Z=120 \cite{ben03} using mean-field
calculations.

A qualitatively different interpretation for the bubble existence is
invoking a specific microscopic quantum mechanical effect: the s-states
being the only wave functions with non zero value at r=0, the depopulation
of this level leads to a depletion of the density in the center of the
nucleus. Such an interior-peaked shape of the 3s wave function has been
measured using electron scattering on $^{206}$Pb and $^{205}$Tl
\cite{cav82}. However, the depletion in the interior of these nuclei remains
small since in heavy nuclei the 2s$_{1/2}$ state is already filled and
contributes to the density at r=0. In an early Hartree-Fock+BCS calculation
a strong depletion in the interior of $^{36}$Ar and $^{200}$Hg nuclei was
predicted \cite{cam73}, which the authors defined as a bubble occurrence.
However, their effective nucleon-nucleon interaction was not considered as
sound, and it was shown that using more elaborated interactions prevent
bubble formation in $^{36}$Ar \cite{bei73}. It should be noted that the key
point for the bubble formation in the s-d region is the inversion of the
proton 2s$_{1/2}$ and 1d$_{3/2}$ states with respect to the standard shell
model.

The emergence of experimental studies on exotic nuclei gathered numerous
signals of shell structure modification these last decades (see e.g.
\cite{oza00,bec06,ele07}). It is therefore legitimate to suspect that in
more neutron-rich nuclei, the s and d state crossing may occur. Such an
inversion was recently predicted in the $^{46}$Ar nucleus using relativistic
mean field (RMF) calculations. A strong depletion in the center of the
proton density was obtained \cite{tod04}. However pairing effects could
preclude the bubble effect due to the occupancy of 2s$_{1/2}$ state. Such a
study has been extended to N=28 isotones \cite{pie07}. The mechanism of this
inversion in neutron-rich nuclei has been analyzed in Ref. \cite{gra07}, and
it is due to the conjunction of several factors: the spin-orbit potential
modification, the tensor force, as well as the lowering of the proton
potential and the extension of the proton density due to the presence of a
neutron skin. Predictions for more neutron-rich nuclei show an increase of
the gap between the inverted s and d states, strongly supporting the
formation of proton bubbles in $^{60-68}$Ar.

In this paper we analyze the general features of bubble occurrence in exotic
nuclei, on an illustrative case. The bubble occurrence is a direct
consequence of the s$_{1/2}$ depopulation, which could be obtained either by
lowering the proton number, or by a level inversion involving an s state.
This work is devoted to the latter case. The possible candidates are
reviewed in Section II, showing that proton bubbles in very neutron-rich Argon
isotopes are the most likely. Several experimental methods to detect such a
bubble are then investigated : charge density measurements (Section III) and
direct reactions (Section IV).

\section{Candidate nuclei}

\subsection{Possible candidates}

\begin{figure}
\begin{center}
\epsfig{file=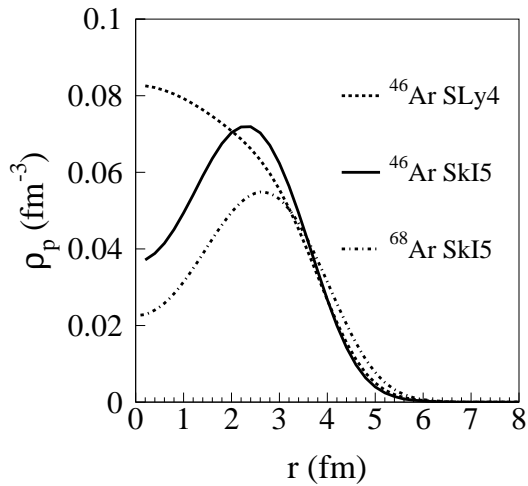,width=8.5cm}
\end{center}
\caption{Proton densities of $^{46}$Ar calculated with the SkI5 interaction
(full lines) and SLy4 \cite{cha98} (dashed lines) in the HF approach. The $^{68}$Ar proton density
calculated with the SkI5 interaction is also shown (dashed-dotted lines)}
\label{fig1}
\end{figure}

Bubble candidates are nuclei where the s state is depopulated. In some cases
this occurs due to an inversion between the s state and the one usually
located above. Therefore, either proton or neutron states inversion between
(2s$_{1/2}$,1d$_{3/2}$) states or (3s$_{1/2}$,1h$_{11/2}$) states are
possible candidates. In the proton bubble case, these inversions correspond
to Ar and Hg isotopes. In the neutron bubble case, they correspond to the
N=18 and N=80 isotones. A first experimental signal for an inversion may be
given by the J$^\pi$ of the ground state of the odd nuclei having one
additional nucleon compared to the bubble nuclei (N or Z equals to 19 or
81). A J$^\pi$=1/2$^+$ together with a spectroscopic factor close to one is
an indication for an s$_{1/2}$ state located above the usual 1d$_{3/2}$ or
1h$_{11/2}$ state. In the case of the Potassium isotopes, such values have
been measured in $^{47}$K \cite{ogi87}, showing that the s-d states are
almost degenerated. An (2s$_{1/2}$,1d$_{3/2}$) inversion is even predicted 
by RMF calculations \cite{tod04}. In Ref. \cite{gra07}, the study along the
Calcium isotopic chain shows that the s-d states inversion is strengthened
for very neutron-rich nuclei (A$\geq$58). Hence the study of very
neutron-rich Argon isotopes such as $^{68}$Ar using the HF method provides an
illustrative study of the bubble manifestation. Fig. \ref{fig1} shows the
proton densities of $^{68}$Ar, calculated with spherical HF model
using the SkI5 \cite{rei95} force. The numerical details are the same as in
Ref. \cite{gra07}. A strong depletion in obtained in the center of the
nucleus, with this SkI5 parametrization. It should be noted that the large
s-d state inversion for very neutron-rich nuclei in the $^{68}$Ar region is
also obtained with several other interactions \cite{gra07} as well as
shell-model calculations \cite{gau07}.

In the case of the N=18 isotones, the measured J$^\pi$ values of the ground
states of the N=19 isotones are J$^\pi$=3/2$^+$ \cite{fir96}, showing no
clear signal for an inversion. On the contrary all the measured J$^\pi$ of
the Z=81 isotopes are 1/2$^+$ in the ground state, showing a systematic
inversion leading to a possible proton bubble in the Hg isotopes. However,
as stated above, the depletion in the center of the proton density of the Hg
isotopes is less pronounced than for the Argon isotopes, since in heavy
nuclei the 2s$_{1/2}$ state is already filled and contributes to the density
at r=0. For the N=81 isotones, J$^\pi$ are measured for only 3 nuclei and
J$^\pi$=1/2$^+$ is found in $^{145}$Gd and $^{147}$Dy \cite{fir96}. In this
case also, the bubble effect in heavy nuclei is expected to be less
important than in intermediate mass nuclei. To summarize, the optimal
manifestation of a bubble could be for protons in neutron-rich Argon
isotopes. We will therefore focus on these nuclei in the following.

It should be noted that deformation can play against a bubble occurrence: the
degeneracy removal attenuates the bubble effect generated by the inoccupancy
of the 2s$_{1/2}$ state. In the case of $^{46}$Ar, several N=28 studies used
deformed mean-field models \cite{wer96,lal99,hil06}, and also the Generator
Coordinate Method (GCM) based on Gogny force \cite{rod02}. In the HFB
calculations, $^{46}$Ar is predicted as a soft nucleus, either spherical or
with a small deformation parameter in the ground state. More neutron-rich
Argon isotopes are also predicted spherical \cite{hil06}. In the GCM
approach, the ground state of $^{46}$Ar is predicted as a possible shape
coexistence state. In the following we will consider the spherical case for
both $^{46}$Ar and $^{68}$Ar nuclei in order to study the bubble hypothesis
in this framework.

\subsection{Pairing effects}

Pairing effects may hinder the bubble formation since scattered pairs could
populate the s$_{1/2}$ state, decreasing the depletion in the center of the
nucleus. The occupancy factor of the s-state due to pairing effects may be
modified. The study of the proton pairing effect in Argon isotopes is a
delicate task : it is occurring in a two-hole state from a doubly magic
nucleus. It is known that the HFB approximation is not well designed to such
a situation, leading to an overestimation of the pairing effects. For
instance, we have performed HFB calculations on $^{46,68}$Ar using the SkI5
interaction and a delta density-dependent pairing interaction, of the form :

\begin{equation}\label{eq:vpair}                                                
V_{pair}=V_0\left[1-\eta\left(\frac{\rho(r)}{\rho_0}\right)^\alpha\right]
\delta\left({\bf r_1}-{\bf r_2}\right)                                                           
\end{equation}  

with $\eta$=1, $\alpha$=0.5 and $\rho_0$=0.16 fm$^3$. The magnitude
V$_0$=-330 MeV.fm$^3$ of the pairing interaction is obtained by reproducing
the two-proton separation energy of $^{46}$Ar. Fig. 2 displays the proton
density of $^{68}$Ar, showing a reduction of the bubble effect, but still a
depletion is predicted in the center of the nucleus.

In $^{46}$Ar the pairing interaction plays an even more dramatic role: the s
state occupation is predicted to be 54\%. The corresponding proton densities,
with and without pairing effects, are displayed on Fig. 1 and 2,
respectively. As expected the 1d$_{3/2}$ state is depopulated, due to the
pairing effect. Shell-model calculations also predict around 50\% occupation
probability for the 2s$_{1/2}$ state \cite{gau07}. However, it should be
noted that the 2s$_{1/2}$ and 1d$_{3/2}$ occupation probabilities have not
been measured yet, leaving a small possibility for a bubble occurrence in
$^{46}$Ar. 

In Ref. \cite{tod04} the splitting of the neutron p-states in $^{46}$Ar is
weakened by the proton density depletion in the center of the nucleus, which
modifies the spin-orbit term in the interior: the proton depletion is
related to the neutron splitting of the p orbits measured in Ref.
\cite{gau06,gau07}. The weakening of the neutron p states splitting with the
occupancy of the 2s1/2 orbital has been recently studied in \cite{pie07}
from Si to Ca N=28 isotones. A proton depletion is predicted for nuclei
having a low 2s$_{1/2}$ occupation probability.

\begin{figure}
\begin{center}
\epsfig{file=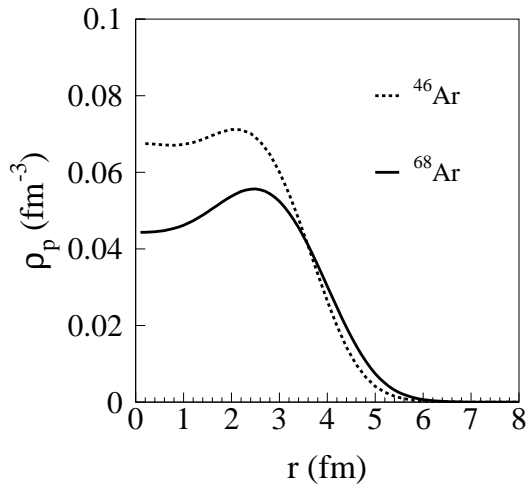,width=8.5cm}
\end{center}
\caption{ Proton densities of $^{46}$Ar (dashed lines) and $^{68}$Ar
calculated with the SkI5 interaction in the HFB approach.} \label{fig2}
\end{figure}

The presence of a proton bubble in $^{46}$Ar is subject to
significant uncertainties, due to the dependence on the pairing and the
Skyrme interactions. Hence, experimental indications of a bubble in this
nucleus are of relevant importance. The probability of a proton bubble
manifestation increases when going from $^{46}$Ar to much more neutron-rich
nuclei such as $^{60-68}$Ar. As mentioned above, several
interactions used in Ref. \cite{gra07} predict an inversion between the
2s$_{1/2}$ and 1d$_{3/2}$ proton states, with a large gap around $^{70}$Ca,
showing that the proton bubble prediction for very neutron-rich Argon
isotopes is a general feature.

In the following we choose $^{46}$Ar as an illustrative example. We will
therefore consider the bubble hypothesis through the HF predictions in
$^{46}$Ar using the SkI5 interaction (Fig. \ref{fig1}). The no-bubble
hypothesis can be obtained either by SLy4-HF (Fig. \ref{fig1}) or by
SkI5-HFB (Fig. \ref{fig2}) calculations. We have checked that the two latter
cases give very similar results: in Ref. \cite{gra07} HF calculations using
SLy4 do not predict any inversion of the s and d states, implying no bubble.
Hence we will describe the no-bubble hypothesis using the HF calculations
with the SLy4 interaction. The case of $^{46}$Ar is chosen only for
illustrative purposes since experimental data are available on this nucleus.
All conclusions below remain valid in the case of $^{68}$Ar.

\section{Probing the charge density}

Although there are strong indications that the 2s$_{1/2}$ is half filled in
$^{46}$Ar, we choose this nucleus to illustrate experimental signals
associated with the bubble occurrence, as explained above. It should be
noted that similar conclusions could be drawn on $^{68}$Ar or other bubble
nuclei with respect to the present calculations. $^{46}$Ar is an unstable
nucleus which can be produced in present exotic beam facilities with a
typical intensity of 10$^3$ pps. Several experimental studies have already
been performed and we will analyze in the following these data with respect
to a bubble manifestation. The most direct observable related to a proton
bubble is the charge density.

\subsection{The charge radius}

Charge radii of Ar isotopes have been precisely measured from $^{32}$Ar to
$^{46}$Ar \cite{bla06,kle96}, using laser spectroscopy of fast beams at the
ISOLDE isotope separator. The obtained value for $^{46}$Ar is R$_c$=3.44
$\pm$ 0.01 fm. However, the calculated $^{46}$Ar r.m.s. charge radii with
the HF or HFB models are mainly sensitive to the surface part of the proton
density, and are not adequate quantities to detect a bubble in the center of
the nucleus: a typical variation of 0.02 fm is found between calculations
assuming the bubble and the no-bubble hypothesis. This difference is not
enough to lead to a clear signature of a bubble occurrence. It is therefore
necessary to look for complementary observables which could be more
sensitive to a bubble manifestation.

\subsection{Electron scattering}

\begin{figure}
\begin{center}
\epsfig{file=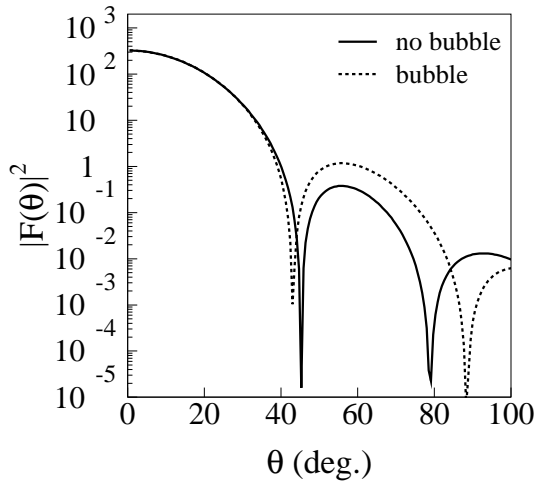,width=8.5cm}
\end{center}
\caption{Angular dependence of the form factor obtained for 300 MeV electron
scattering on $^{46}$Ar, using either the HF-SkI5 density (dashed line) or
the HF-SLy4 density (solid line)} \label{fig3}
\end{figure}

The ideal experiment to probe the charge density would be electron
scattering. Presently electrons scattering on unstable nuclei like $^{46}$Ar
is not possible, but it is expected in next generation facilities such as
RIBF in Riken \cite{mot05} of FAIR in GSI \cite{rub06}. In this case
accelerated electrons would scatter on a radioactive beam of $^{46}$Ar kept
in a storage ring. Such an experiment may be feasible in the next decade. It
is therefore worthwhile to check how accurately the electron scattering
experiment could probe the proton bubble presence. We have performed
calculation for 300 MeV electron scattering $^{46}$Ar. Fig. \ref{fig3} shows
the angular dependence of the form factor defined by :

\begin{equation}
F(q)=\int {\rho_c(r) e^{i\vec{q}.\vec{r}} d\vec{r}}
\end{equation}

where $\rho_c$ is the charge density calculated with HF model using SLy4 or
SkI5 interactions, and q is the transferred momentum, related to the
incident momentum p and the scattering angle $\theta$: 
 
\begin{equation}
q=2p\sin\frac{\theta}{2}
\end{equation}

Both the no-bubble and the bubble cases exhibit a diffraction pattern.
However, above 45 deg. the angular distribution is located at higher
magnitude in the case of a bubble than without a bubble. Around 80 deg. an
angular shift between the two distributions is noticed, the minima of the
bubble distribution being 10 deg. larger than the normal one. At 90 deg. the
two angular distributions are in opposite phases, showing that this angle is
the optimum one to disentangle between the two hypothesis. At larger angle
the angular shift increases but the magnitude becomes too small for a clear
measurement. In summary, an optimal experiment to search for a bubble in
very neutron-rich Argon isotopes would be elastic electron scattering at 300
MeV, focused on scattered electrons between 40 degrees and 100 degrees.

\section{Direct reactions and collective modes}

As stated above, such an experiment with electron scattering on exotic beams
is not feasible yet. It is therefore interesting to analyze what hints
direct reactions could provide with respect to a bubble manifestation. It is
well known that direct reactions probe the surface of the nucleus, so it is
not possible to directly detect a bubble located in the center of the
nucleus. However, in the Argon isotopes, the bubble is a straightforward
consequence of the inversion between the 1d$_{3/2}$ and the 2s$_{1/2}$
levels, provided that the 2s$_{1/2}$ remains depopulated. Direct reactions
are the most accurate tool to study these features. For instance transfer
reactions such as (d,$^3$He) would allow to measure the spectroscopic
factors associated to the 2s$_{1/2}$ level. Collective modes such as giant
resonances and low-lying states may also provide useful information about a
bubble manifestation. We shall again illustrate the bubble occurrence on
$^{46}$Ar, which could more probably occur on more neutron-rich isotopes.

\subsection{Giant resonances}

Giant resonances are collective modes involving all the nucleons. To
disentangle the effect of the bubble from the one due to the interaction,
namely due to the modification of the single particle spectrum, HF and HFB
calculations have also been performed with both the SLy4 and the SkI5
interactions in the $^{48}$Ca case, where no bubble effect is present. 

In order to perform such a study and compare with the data, HFB+QRPA
calculations have been performed. All detailed presentation of the model can
be found in Ref. \cite{kha02}. The residual interaction is derived from the
Skyrme functional, as well as the pairing residual interaction, derived from
the pairing part of the functional used in the HFB calculation. We have
first checked that the HFB+QRPA results using SkI5 give results similar to
the HF+RPA using SLy4, that is the no-bubble hypothesis. Therefore it is
sufficient to perform calculations using SLy4 without pairing effects. The
Landau-Migdal approximation is used, leading to a small breaking of the
self-consistency, cured by a renormalization of the residual interaction to
set the center of mass spurious mode to zero energy \cite{kha02}. The
typical renormalization values are 0.9.

\begin{figure}
\begin{center}
\epsfig{file=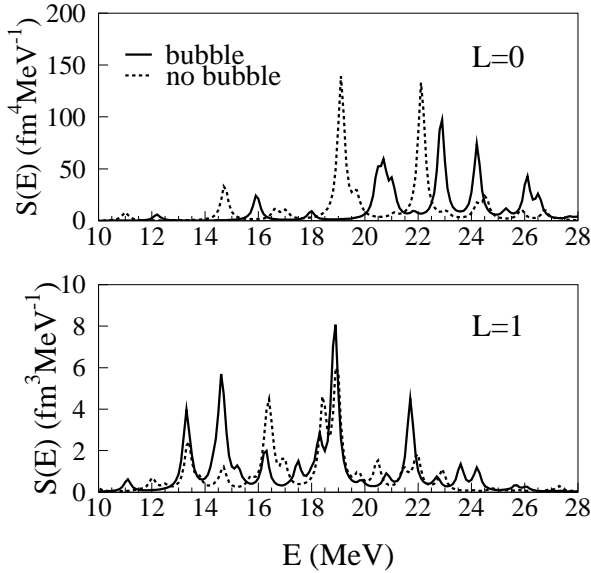,width=8.5cm}
\end{center}
\caption{Isoscalar monopole and isovector dipole strength functions obtained
with the HF+RPA approach and the SkI5 interaction (solid line) or the SLy4
interaction (dashed line) in  $^{46}$Ar} \label{fig4}
\end{figure}

The main effects on giant resonances of $^{46}$Ar which are not present in
$^{48}$Ca are obtained in the low energy part of the dipole and monopole
resonances (Fig \ref{fig4}). No strong discrepancy is observed between the
bubble and the normal cases in this low energy area. In the case of the
monopole response, the lowest energy state is shifted by 1.5 MeV at higher
energy in the bubble case, from 14.5 MeV, to 16 MeV. It may be interpreted
as a slight increase of the nucleus incompressibility for a soft mode in the
bubble case: protons are depleted in the center of the nucleus and are
redistributed in the surface, making more difficult a soft compression of
the nucleus.

The isovector E1 strength exhibits an increase of the low-lying part, around
13 MeV, in the bubble case. This effect is not present in the SkI5
calculations on $^{48}$Ca: the strength enhancement is more important in the
$^{46}$Ar case than in the $^{48}$Ca case. This soft mode may be interpreted
as the oscillation of neutron with respect to protons which are surface
localized.

However, the above mentioned effects are hardly sizable, especially for
giant resonances related effects, where the spreading width implies a large
experimental width which can mask these effects.

\subsection{Low-lying states}

The low-lying states of $^{46}$Ar have been recently studied using direct
reactions, either by Coulomb excitation \cite{gad03} or by proton
scattering in inverse kinematics \cite{ril05}. As stated above, they are
an accurate tool to study the level inversion as well as the eventual
population of the 2s$_{1/2}$ state due to the pairing effect. 
 
\begin{figure}
\begin{center}
\epsfig{file=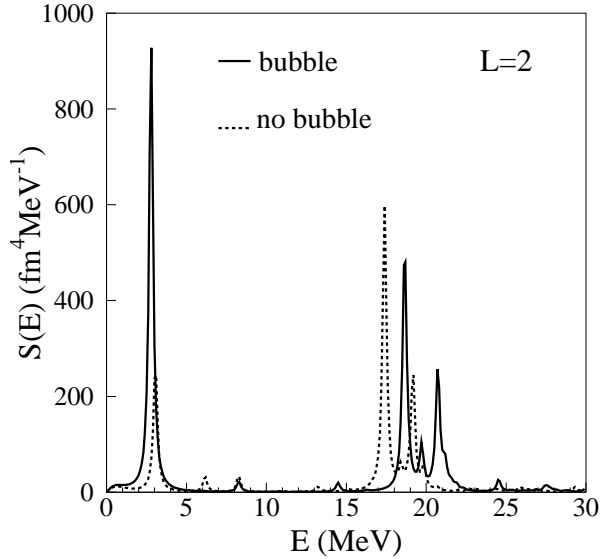,width=8.5cm}
\end{center}
\caption{ Isoscalar quadrupole strength function
obtained with the HF+RPA approach and the SkI5 interaction (solid line) or
the SLy4 interaction (dashed lines) in $^{46}$Ar} \label{fig5}
\end{figure}

The study of the first 2$^+$ state of $^{46}$Ar by proton scattering shows a
significant contribution of the neutrons to the excitation \cite{ril05}.
Hence this state is not the best probe for the proton shell structure.
However, the study of the proton contribution through the B(E2) value could
provide relevant information. Fig. \ref{fig5} shows the isoscalar quadrupole
response of $^{46}$Ar, calculated by HF+RPA with the SLy4 interaction (no
bubble) and the SkI5 one (bubble). There is a strong enhancement of the
strength of the 2$^+_1$ state in the bubble case. The predicted B(E2) values
strongly differs: 24 e$^2$.fm$^4$ in the normal case and 256 e$^2$.fm$^4$ in
the bubble case. This last case is in very good agreement with the data
\cite{gad03} : B(E2)=218 $\pm$ 31 e$^2$.fm$^4$. This enhancement is partly
due to the inversion between the 1d$_{3/2}$ and 2s$_{1/2}$ states: more
proton particle-hole configurations contribute in the bubble case than in
the normal case. However, the main effect comes from the RPA residual
interaction, which is very different in the two cases for this low-lying
mode. Namely, the density-dependent terms of the residual interaction could
explain this change, since the density profile is different in the two
cases. Moreover as stated above, HFB+QRPA calculations using the SkI5
interaction, give similar results to the normal case, showing that it is the
density dependence of the residual interaction which drives the B(E2)
enhancement. These terms are generated by the density-dependent part of the
effective nucleon-nucleon interaction itself. Therefore the study of bubble
nuclei could not only provide information on the spin-orbit term as stated
in Ref. \cite{tod04,pie07}, but also on the importance of the
density-dependent term of the nucleon-nucleon interaction. The measurement
of the B(E2) in bubble nuclei may open a way to explore the impact of this
term in the energy density functional. This shows that the manifestation of
a bubble in $^{46}$Ar might be necessary to reproduce the data, namely the
first 2$^+$ state.

\section{Conclusions}

We have investigated the bubble occurrence in nuclei, due to an inversion of
the s$_{1/2}$ state with the state usually located above. The neutron-rich
Argon isotopes are the best candidates of the nuclear chart, since the
proton 2s$_{1/2}$ state largely contributes to the density in the center of
the nucleus, and because the inversion effect is known to increase with the
neutron number. Very neutron-rich nuclei such as $^{68}$Ar are optimal
candidates. In less neutron-rich nuclei such as $^{46}$Ar, the situation is
less favorable since the (2s$_{1/2}$,1d$_{3/2}$) inversion is reduced,
implying a significant occupation probability for the 2s$_{1/2}$ state,
leading to a weakening of the bubble effect. 

$^{46}$Ar is chosen as an illustrative case, considering both the bubble and
no-bubble hypothesis: experimental data are available on this nucleus, and the
predictions are similar for more neutron-rich isotopes. The p-states neutron
splitting may be related to a proton depletion, and the B(E2) is also very
well reproduced in $^{46}$Ar, considering the bubble hypothesis. These
indications call for a more detailed experimental investigation in this
nucleus, as well as more neutron-rich ones, which could be undertaken by
elastic electron scattering with the next generation exotic beam facilities
such as FAIR or RIBF.

The bubble effect is weakened by pairing effects which can populate the
2s$_{1/2}$ state. However the inverted s and d level spacing increases for
very neutron-rich Argon nuclei, and even in the case of a strong pairing,
some isotopes more neutron-rich than $^{46}$Ar should exhibit a bubble, as
predicted in the dramatic case of $^{68}$Ar. Hence, direct reactions on
$^{60,62}$Ar in next generation radioactive beam facilities will also be of
great interest.  They should focus on low-lying states since giant
resonances do not exhibit strongly different patterns in the case of a
bubble, compared to the normal case.

The discovery of bubbles in nuclei is certainly an important issue for nuclear
structure. This study calls for other theoretical investigations, as well as
an important experimental program devoted to frontline exotic nuclei, such
as neutron-rich argon isotopes. 

\vskip 0.5cm

\noindent{\bf Acknowledgments} The authors thank F. Leblanc, O. Sorlin and
D. Verney for fruitful discussions.

\newpage

\end{document}